\documentclass[a4paper,11pt]{article}

\usepackage{jheppub} 
\usepackage{graphicx, graphics, color}
\usepackage[T1]{fontenc} 

\usepackage{amsmath, amssymb, bm, graphicx, graphics, color, mathrsfs}
\newcommand{\red}[1]{\textcolor[rgb]{1.00,0.00,0.00}{#1}}

\newcommand{\be}{\begin{equation}}
\newcommand{\ee}{\end{equation}}
\newcommand{\ben}{\begin{eqnarray}}
\newcommand{\een}{\end{eqnarray}}

\newcommand{\cO}{{\cal O}}

\newcommand{\cM}{{\cal M}}
\newcommand{\cR}{{\cal R}}

\newcommand{\p}{\partial}
\newcommand{\na}{\nabla}

\newcommand{\ep}{\epsilon}

\title{\boldmath 
Holographic vortices in the presence of dark matter sector }

\author[1]{Marek Rogatko\note{rogat@kft.umcs.lublin.pl, marek.rogatko@poczta.umcs.lublin.pl}}
\author[2]{Karol I. Wysokinski\note{karol.wysokinski@umcs.pl}}
\affiliation{Institute of Physics \\
Maria Curie-Sk{\l}odowska University \\
20-031 Lublin, pl. Marii Curie-Sk{\l}odowskiej 1, Poland}

\abstract{
The {\it dark matter} seem to be an inevitable ingredient of the
total matter configuration in the Universe and the knowledge how the {\it dark matter} affects the properties 
of superconductors is of vital importance for the experiments aimed at its direct detection. 
The homogeneous magnetic field acting perpendicularly  to the surface of (2+1) dimensional 
s-wave holographic superconductor  in the theory with {\it dark matter} sector 
has been  modeled by the additional $U(1)$-gauge field representing dark matter and coupled 
to the Maxwell one. As expected the free energy 
for the vortex configuration turns out to be negative. Importantly its value is lower in the presence  
of {\it dark matter} sector. This feature can explain why in the Early Universe
first the web of {\it dark matter}  appeared and next on these gratings the 
ordinary matter forming cluster of galaxies has formed.}

\keywords{Gauge-gravity correspondence,
Holography and condensed matter physics (AdS/CMT), Black Holes}

\begin{document} 

\maketitle
\flushbottom


\section{Introduction}
\label{sec:intro}
Proposed as the equivalence between type IIB superstring theory on $AdS_5 \times S^5$ spacetime and ${\cal N}=4~ SU(N)$
supersymmetric Yang Mills theory on $(3+1)$-dimensional boundary \cite{mal}-\cite{gub98} and generalized to  the other
gravitational backgrounds \cite{rev1}, the gauge/gravity duality helps us to understand the physics of strongly correlated systems.

The main feature of the string theories is that gravity emerges from them in a natural way. In this context the string theories
dominated by the classic gravity configurations provide a basis for the 
weak-strong duality \cite{sac12}. In technical terms \cite{gre13} it means that
the partition function of the field theory $Z[A({\bf x},t)]$, where $A({\bf x},t)$ is the source field that couples
to the currents ${\bf j}({\bf x},t)$, equals to that of the gravity dual 
$Z[A({\bf x},r,t)]$ with the boundary condition $\lim_{r\rightarrow \infty}A({\bf x},r,t) =A({\bf x},t)$. The latter partition function 
is prevailed by the classical configurations and thus given by the exponential of the classical action of the aforementioned configurations.

The AdS/CFT correspondence has been successfully applied for the description of the superconducting 
phase transition of the single s-wave superconductor \cite{har08}, as well as, the
other symmetries  \cite{che10}-\cite{cai15}.
The effects of the higher order curvatures \cite{pan11}-\cite{cai11back} of the gravity background and non-linearities in the electrodynamic theory \cite{zha15}-\cite{jin12}
have been widely studied. 

The aforementioned studies were also extended by the modification of gravity theory by considering
the five-dimensional AdS solitonic metric \cite{hor98}, which enables to construct a model of the holographic
insulator/superconductor phase transition at zero temperature \cite{nis10}. The AdS soliton line element
dual to a confined field theory with a mass gap, imitates an insulator phase \cite{wit98a}. 

The effects of magnetic field (which will be crucial in our further studies) on holographic superconductors were also intensively studied.
It was revealed that by adding magnetic charge to a black hole, 
the holographic superconductor could be immersed in magnetic field \cite{alb08}.
The studies of the problem in question were also conducted in Gauss Bonnet gravity and in Weyl-corrected 
and non-linear electrodynamics \cite{ge10}-\cite{roy12}.
Holographic vortices and droplets influenced by magnetic field were elaborated in \cite{alb09}-\cite{roy13}, while the vector condensation generated by this factor was studied in \cite{cai11b,cai13}.
On the other hand, the magnetic penetration depth was computed in \cite{har08}, where
it was envisaged that the holographic superconductors are of type II.
In order to find the superconducting coherence length and magnetic penetration depth\cite{mae08} one perturbs the AdS-Schwarzschild 
system mimicking s-wave superconductor near the critical temperature.
These works triggered the  investigations in p-wave as well as 
d-wave holographic superconductors \cite{zen10a}-\cite{zen10b}.
It has been found that the superconducting coherence length was proportional to $(1- T/T_c)^{-1/2}$ near 
the critical temperature. Similar dependence of the magnetic penetration depth
 $\lambda \propto (1- T/T_c)^{-1/2}$ has been observed. These results agree with the standard Ginzburg-Landau theory.

In \cite{roy15} the interesting case of the Abelian Chern-Simons Higgs model in $(2+1)$-dimensions was elaborated. From this point of view
the higher derivative corrections to the Abelian gauge sector in the bulk were studied \cite{roy14}. The vortex lattice 
solution in a holographic superconductors constructed
from a charged scalar condensate was elaborated in \cite{mae10}. The  perturbative solution near 
the second-order phase transition represents 
the holographic Abrikosov lattice. On the other hand, the holographic vortex-flow solution was given in \cite{mae11}.

The other crucial problem is the question about the possible matter configuration in the AdS spacetime.
This problem was considered in the case of strictly stationary Einstein-Maxwell AdS spacetime \cite{shi12}
and its generalization to the low-energy limit of the heterotic string theory with arbitrary number
of $U(1)$-gauge fields \cite{bak13}. It was concluded that the spacetimes in question can not allow for the
existence of nontrivial configurations of complex scalar or form fields.

The origin of cosmic structures can be understood in terms of evolution of matter perturbations arising after 
inflationary period. The 
{\it dark matter} constitutes the key ingredient in the aforementioned processes, 
as a potential wells where the ordinary matter accrete. If {\it dark matter} appears in the form of particles
coupled to ordinary matter then a certain level of emitted radiation is expected. 
The non-gravitational signals are anticipated from interaction of dark matter with ambient medium, annihilation or decay
by a direct emission or through particle production. The non-gravitational signals of it should be 
in principle proportional to the density of clumps of dark matter. Recently,
a new method to constrain the process of elastic scattering between dark matter and the Standard Model 
particles in the Early Universe was proposed \cite{reg15,ali15}.

{\it Dark matter} may also leave its footprints during collapse of neutron stars and in the 
first generation of stars \cite{bra14}-\cite{lop14}. On the other hand, the behavior of {\it dark matter}
and dark energy collapse is studied numerically from the point of view of possible features of black hole 
and wormhole formations \cite{nak12}-\cite{nak15a}. This is an important problems in studies of the black hole growth
in the Early Universe.

The dark matter non-gravitational interactions, having implications for the particle physics 
beyond the Standard Model, triggered search for high-energy
photons produced by annihilation which in turn causes the resurgence in inspection of gamma-rays 
emission from the dwarf galaxies \cite{ger15}. This problem is also interesting in the
context of planing varieties of gamma-rays telescopes in the range of energy from 1 to $100 MeV$ \cite{bod15}. 
Investigations of dilatonic like coupling to photons which can be caused
by ultralight dark matter was reported in \cite{til15}. This process can induce oscillations in the fine-structure constant.

The {\it dark matter} sector model may have
a strong support by some astrophysical data such as observation of $511$ keV gamma rays \cite{integral} 
experiments detecting the electron positron excess in galaxy
\cite{atic, pamela}, and  possible explanation of muon anomalous magnetic moment\cite{muon}.
Recent astronomical observations reveal that
collisions among galaxy clusters provide new tests of non-gravitational forces 
acting on dark matter \cite{massey15a,massey15b}. The experiments in question may disfavor some 
extensions of the Standard Model. 

The idea that {\it dark matter} sector is charged under a new $U(1)$-gauge group and is coupled to the ordinary Maxwell
field gained experimental verifications. The low-energy $e^+~e^-$ colliders offer 
the environment to probe low-energy mass {\it dark sector} \cite{babar14}.
The data from BABAR detector were revealed bounded with a search for dark photon production in the range 
of mass $0.02 < m < 10.2 GeV$, but no significant
signal has been observed. The other experiments are planned to cover the energy region $15 \leq m \leq 30 MeV$.

It is thus obvious that study of {\it dark matter} and its observational consequences is both timely and
of  crucial importance. In this paper we add a novel aspect to it by analyzing the behavior of holographic 
superconductors.

We shall try to answer the question how a {\it dark matter} sector modifies the ordinary phase 
transitions known from the pervious studies and what the role of
the dark matter coupling constant $\alpha$, binding dark matter fields with ordinary Maxwell gauge field is.
We investigate analytically the perturbation of the dual gravity theory to reveal 
the superconducting coherence length and magnetic penetration depth close to the  
superconducting phase transition point. The main aim of this paper is to study the influence of the 
{\it dark matter} sector on the  holographic vortex. In the previous works \cite{nak14}-\cite{pen15}, 
various aspects of phase transitions in s-wave and p-wave holographic superconductor theory 
with ordinary matter sector coupled to another $U(1)$-gauge field which describes {\it dark matter} have been elaborated. 
One may expect that {\it dark matter} authorizes a part of a larger particle sector interacting with 
the visible matter  and not completely decoupled \cite{vac91}-\cite{dav13}.

The case of homogeneous magnetic field acting perpendicularly 
to the surface of s-wave holographic superconductor 
in the theory with {\it dark matter} sector has been considered.
{\it Dark matter} has been modeled by the additional $U(1)$-gauge field coupled to Maxwell one
and $(2+1)$-dimensional vortices have been studied in the considered theory. 
Free energy for the vortex configuration turns out to be negative and it is less for 
the configuration with {\it dark matter} sector  than for ordinary matter case. 
One concludes that the dark matter sector vortex configurations are more stable than trivial 
configuration where there is no charge scalar condensate. On the other hand,
because of the fact that {\it dark matter} vortices build less free energy configuration than ordinary matter ones, 
this fact can explain why in the Early Universe
first form the web of {\it dark matter} and next on these tendrils the ordinary matter condenses forming cluster of galaxies.

The paper is organized as follows. In section 2 we describe the main features of the theory with {\it dark matter} 
section which is mimicked by the coupling between ordinary Maxwell and $U(1)$-gauge fields.
In section 3, to the leading order, we  find perturbatively solutions of the equations of motion in the underlying theory. 
Section 4 is devoted to free energy for the considered system which determines thermodynamics
stability of the {\it dark matter} sector vortex. Section 5 is connected with calculations of the superconducting coherence 
length and magnetic penetration depth which depend on
the $\alpha$-coupling constant of the {\it dark matter} part of the theory. In section 6 we find the analogy of 
the London equation for the holographic vortex in the theory in question
and calculate   the superfluid density  and magnetic penetration depth. Both quantities are affected 
by the influence of the {\it dark matter} section of the considered theory.

\section{Superconductor with dark matter sector in AdS/CFT theory}
\label{sec:vortex}

The gravitational action in (3+1) dimensions is taken in the form
\be
S_{g} = \int \sqrt{-g}~ d^4 x~  \bigg( R - 2\Lambda\bigg),
\label{s_grav} 
\ee
where $\Lambda = - 3/ L^2$ stands for the cosmological
constant, while $L$ is the radius of the AdS spacetime. We shall examine the Abelian-Higgs sector coupled to the 
second $U(1)$-gauge field which mimics the {\it dark matter} sector of the underlying theory \cite{bri11}.  The adequate action incorporating {\it dark matter} is provided by 
\ben
\label{s_matter}
S_{m} = \int \sqrt{-g}~ d^4x  \bigg( 
&-& \frac{1}{4}F_{\mu \nu} F^{\mu \nu} - \left [ \nabla_{\mu} \psi - 
i q A_{\mu} \psi \right ]^{\dagger} \left [ \nabla^{\mu} \psi - i q A^{\mu} \psi  \right ]
- m^2 |\psi|^2 \\ \nonumber
&-& \frac{1}{4} B_{\mu \nu} B^{\mu \nu} - \frac{\alpha}{4} F_{\mu \nu} B^{\mu \nu}
\bigg), 
\een  
where
$F_{\mu \nu} = 2 \nabla_{[ \mu} A_{\nu ]}$ stands for the ordinary Maxwell field strength tensor, while
the second $U(1)$-gauge field $B_{\mu \nu}$ is given by $B_{\mu \nu} = 2 \nabla_{[ \mu} B_{\nu ]}$. 
Moreover, $m,~ q$ represent, respectively a mass and a charge related to the scalar 
field $\psi$. $\alpha$ is a coupling constant between $U(1)$ fields. Varying the action with respect 
to the considered fields one obtains the equations of motion which can be written in the form as
\ben \label{eqm1}
\bigg( \na_\mu - i~q A_\mu \bigg) \bigg( \na^\mu - i ~qA^\mu \bigg) \psi - m^2~\psi &=& 0,\\ 
\label{eqm2}
\tilde \alpha~\na_\mu F^{\mu \nu} &=& j^\nu,
\een
where the current $j^\nu$ yields
\be
j^\nu =  i ~\bigg[ \psi^\dagger~\bigg( \na^\nu - i ~q A^\nu \bigg)~ \psi 
- \psi~\bigg( \na^\nu + i ~q A^\nu \bigg) \psi^\dagger \bigg] = 0,
\ee
and $\tilde \alpha = 1- \alpha^2/4$. In the above relation we have used the equation determining $B_{\mu \nu}$ field
\be
\na_\mu B^{\mu \nu} + \frac{\alpha}{2}~\na_\mu F^{\mu \nu} = 0,
\label{bb}
\ee
in order to eliminate this field and obtain the relation (\ref{eqm2}). 

The presence of the {\it dark matter} sector is marked by the appearance of $\tilde \alpha$ in the equations of motion.
From this point of view the holographic $s$ and $p$-wave superconductor phase transitions 
were previously studied \cite{nak14,nak15,nak15a1} in the presence of dark matter. 
It was revealed that there were some imprints of 
{\it dark matter} sector on  holographic phase transitions. In particular  it has been found \cite{nak15a1} that in the probe
limit the prefactor $\Gamma$ between the value of the  condensing operator in the relation 
$<\cO>=\Gamma \sqrt{1-T/T_c}$ does depend on $\alpha$. $\Gamma \propto \sqrt{\tilde{\alpha}}$.
Neither the superconducting  transition temperature nor critical value 
of the chemical potential for insulator to metal transition depend on $\alpha$.   Much more severe
dependence on the {\it dark matter} coupling has been observed in the backreacting theory \cite{nak14}, where the 
transition temperature depends on $\alpha$. The p-wave superconductors described by the SU(2) Yang-Mills  theory
features {\it inter alia} the increase \cite{rog15} of the transition temperature with increase of the coupling $\alpha$.
On the other hand, the critical chemical potential $\mu_c$ for the quantum phase transition between insulator 
and a metal is a decreasing function of $\alpha$. 

In the theory with the action given by equations (\ref{s_grav}) and (\ref{s_matter}) the gravitational background 
will be provided by the four-dimensional AdS-Schwarzschild black hole background  of the form
\be
ds^2 = - f(r) dt^2 + \frac{1}{f(r)}dr^2 + \frac{r^2}{L^2}~(dx^2 + d y^2),
\ee
where the metric function reads
\be
f(r)=\frac{r^2}{L^2}\left(1-\frac{r_+^3}{r^3}\right).
\ee
L is the radius of the AdS spacetime and $r_+$ is the horizon of the black hole.
The Hawking temperature of the black hole is found to be
\be
T=\frac{3r_+}{4\pi L^2}.
\ee
In what follows, without loss of generality, we shall
set the spacetime radius $L$ and the charge related to the scalar field equal to 1. One also changes the variables
replacing the space coordinate $r$ by $u=r_+/r$. This replacement results in the
line element of the form
\be
ds^2 = - f(u) dt^2 + \frac{r_+^2}{u^4 ~f(u)}du^2 + \frac{r_+^2}{u^2}~(dx^2 + d y^2),
\ee
where $f(u) = r_+^2/u^2~ (1-u^3)$. 

We are interested in the properties of superconductors subject to the external magnetic
field. As it is well known from experiments on superconducting systems and in agreement with Ginzburg-Landau
theory, type I superconductors are characterized by the single value of the critical magnetic field $B_c$, beyond
which the superconductivity is destroyed. Such systems are characterized by the penetration depth $\lambda$
smaller than the superconducting coherence length $\xi$. The existence of the border between normal
metal and superconductor increases its energy and makes the system thermodynamically unstable. On the other hand,
in the opposite limit of $\lambda/\xi>>1$ (type II superconductors) the existence of the border decreases the energy and
allows the magnetic field to penetrate into the metal in the form of vortices. The process starts at the 
"lower magnetic field" $B_{c1}$ and continues until ''upper magnetic field`` $B_{c2}$, when the vortices 
are so dense that they overlap and the system undergoes superconductor to normal metal transition.  
The vortices are normal systems at their cores. We do not consider here the dynamics of the gauge field.
It is important in analysis of various properties of superconductors \cite{dome10,mont12} in particular the Meissner effect.

One should remark that in the considered theory there are in principle two ways of choosing the magnetic field. 
One is the ordinary magnetic field introduced by the appropriate component of the $U(1)$-gauge Maxwell field, $A_y$.
The other, is to choose the magnetic field of {\it dark matter} sector and then by the equation of motion (\ref{bb}) receive 
the Maxwell component of $A_y$ which enters the relation (\ref{eqm2}).
Choosing the second possibility  we assume constant dark matter
magnetic field $\hat B$
\be
B_y = \hat B~x.
\ee
Thus, having in mind equation (\ref{bb}), we obtain
\be
A_y = \frac{2}{\alpha}~\bigg( c_1 - \hat B \bigg)~x +c_2,
\label{mag}
\ee
where $c_{1,2}$ are arbitrary constants. 
In our further considerations we shall keep the general form of $A_y$ and in the resulting conclusion one specifies the influence of the two aforementioned possibilities of choosing magnetic field.

To proceed further we shall solve the equations of motion perturbatively in the probe limit. Namely, 
we define the deviation parameter $\ep_H = (B_{c2} -B)/B_{c2}$, with the property that $\mid \ep_H \mid \ll 1$, 
and expand gauge fields $A_{\mu}({\bf x},~u)$, the scalar field $\psi ({\bf x},~u)$ 
and the current $j_{\mu}({\bf x},~u)$ in series of $\ep_H$ 
\ben \label{aa1}
A_{\mu}({\bf x},~u) &=& A_{\mu}^{(0)}({\bf x},~u) + \ep_H~A_{\mu}^{(1)} ({\bf x},~u) + \dots,\\ \label{aa2}
\psi ({\bf x},~u) &=& \ep_H^{\frac{1}{2}}~\psi_1({\bf x},~u)  + \ep_H^{\frac{3}{2}}~\psi_2({\bf x},~u)  + \dots, \\ \label{aa3}
j_{\mu}({\bf x},~u) &=& \ep_H~j^{(1)}_{\mu}({\bf x},~u) + \ep_H^2~j^{(2)}_{\mu}({\bf x},~u)  + \dots,
\een
where $\bf x = (x,~y)$. 
The above chosen expansions are analogous to that used previously \cite{mae08,zen10a,zen10b,roy15,roy14,mae10,mae11}.
The different powers of  $\ep_H $ are related to the equations of motion for the scalar field $\psi$ and
$A_t$ as discussed in \cite{mae08}. They envisage the fact that the scalar field $\psi$ acts as an order parameter 
and vanishes at the critical value of the magnetic field in the mean field like manner.
In what follows we solve and analyze the solutions of the equations (\ref{eqm1}) and (\ref{eqm2}), for $A_\mu$ and $\psi$, up to the leading order.

\section{Leading orders of the equations of motion}
\label{sec:zero}

\subsection{Zeroth order}
Substituting the relations (\ref{aa1})-(\ref{aa3}) into (\ref{eqm1}) and (\ref{eqm2}), we obtain  
the zeroth order equations of motion
\be
\tilde \alpha~\na_\nu F^{\nu \mu (0)}  = 0.
\ee
In accordance with the assumed dependence of the fields in question on the parameter $\ep_H$, in zeroth order only
the components of the gauge fields have non-zero values.
In order to find the solutions we consider the ansatz as follows:
\be
A_{\mu}^{(0)}({\bf x},~u) = (A_t({\bf x},~u),A_u({\bf x},~u),A_x({\bf x},~u),A_y({\bf x},~u)) =(\varphi(u),~0,~0,~A_y^{(0)}).
\ee
The zeroth order solution generates the chemical potential $\mu$ and the critical magnetic field $B_{c2}$ as given by the relations
\be
A_{t}^{(0)} = \mu~(1 - u), \qquad A_u^{(0)}, \qquad A_{x}^{(0)} = 0, \qquad A_{y}^{(0)} = B_{c2}~x.
\ee
The solution agrees with the fact that at $B=B_{c2}$ the superconducting order parameter and the currents vanish.

\subsection{Next to zeroth order}
In this subsection one derives equations of motion for the scalar field $\psi$,  the Maxwell 
gauge potentials $A_\mu$ and the current. As it follows from the assumed expansion, the non-zero value
of the condensate $\psi_1$ is a coefficient in front of the $\sqrt{\ep_H}$ while the current being proportional to ''square`` 
of the scalar field appears in the order $\ep_H$. This provides a posteriori justification of the expansion (\ref{aa1})-(\ref{aa3}). 

\subsubsection{Scalar field equation}
Up to the $\sqrt{\ep_H}$-order the equation of motion for the scalar field can be written as
\be
\p_u \bigg[ f(u)~\p_u \psi_1 \bigg] + \frac{1}{u^2}~\bigg( \Delta - 2i~A_y^{(0)}~\p_y - (A_y^{(0)})^2 \bigg) \psi_1 + \frac{r_+^2~\varphi^2(u)}{f(u)~u^4} \psi_1
- \frac{m^2~r_+^2}{u^4} \psi_1 = 0.
\ee
In order to solve the above equations one separates the variables  
\be
\psi_1(u,~{\bf x}) = \rho_{0}(u)~H({\bf x}) = \rho_{0}(u)~e^{ik_y y} X(x).
\ee
Inserting the above ansatz for $\psi_1(u,~{\bf x}) $ we arrive at the set of equations provided by
\ben
\rho''_{0}(u) &+& \frac{f'(u)}{f(u)}~\rho'_{0}(u) + \frac{r_+^2~\varphi^2(u)}{f(u)~u^4} ~\rho_{0}(u) -  
\frac{m^2~r_+^2}{u^4} ~\rho_{0}(u)  = \frac{\rho_{0}(u) }{\zeta^2~u^2~f(u)},\\
- X''(x) &+& (A_y^{(0)})^2 ~\bigg( x - \frac{k_y}{A_y^{(0)}} \bigg)^2~X(x) = \frac{X(x)}{\zeta^2},
\een
where by $\zeta^2$ we denoted a separation constant.

\subsubsection{Solution for $A_\mu$}
With the choice of gauge in the form $A_u =0$, and assuming that the scalar field is the real one,
the generalized Maxwell type equations at first order in $\ep_H$ are provided by
\ben \label{sa1}
D^{(t)} A^{(1)}_t &=& 2~\frac{A^{(0)}_t ~\mid \psi_1 \mid^2~r_+^2}{\tilde \alpha~u^2},\\ \label{sa2}
D^{(i)} A^{(1)}_m - \p_m \bigg( \delta^{fk}~\p_f A^{(1)}_k \bigg) &=& \frac{j^{(1)}_m~r_+^2}{\tilde \alpha~u^2},\\ \label{sa3}
\p_u \bigg( \delta^{ac}~\p_a~A^{(1)}_c \bigg) &=& 0,
\een
where the differential operators are denoted by
\be
D^{(t)} = u^2~f(u)~\p^2_u + \Delta, \qquad D^{(i)} = \p_u \bigg( u^2~f(u)~\p_u \bigg) + \Delta.
\ee
By $\Delta$ we denoted the two-dimensional Laplacian, $\Delta = \p^2_x + \p^2_y$. As mentioned in \cite{mae10}, 
beside the gauge transformation equaling $u$-component of $U(1)$-gauge field to zero also  there exist 
transformation of the form  $A_{m} \rightarrow A_m - \p_m K({\bf x})$ (the so-called residual transformation).
It can be seen that $<J_\mu> \propto F_{u \mu}$ is invariant under the transformation in question.
This transformation allows us to set $ \delta^{ac}~\p_a~A^{(1)}_c =0$ and get the following equations 
instead of (\ref{sa1})-(\ref{sa3}) \cite{mae10}. Namely, one obtains
\ben
D^{(t)} A^{(1)}_t &=& 2~\frac{r_+^2~A^{(0)}_t }{\tilde \alpha~u^2} ~\rho_0^2(u)~\sigma({\bf x}),\\
D^{(i)} A^{(1)}_x &=& \frac{r_+^2}{\tilde \alpha~u^2} ~\rho_0^2(u)~\ep_x{}{}^{y} ~\p_y \sigma({\bf x}),\\
D^{(i)} A^{(1)}_y &=& \frac{r_+^2}{\tilde \alpha~u^2} ~\rho_0^2(u)~\ep_y{}{}^{x} ~\p_x \sigma({\bf x}),
\een
where $\ep_{ab}= \ep^a{}{}_b$ is the two-dimensional totally antisymmetric symbol with the properties $\ep_{xy} = - \ep_{yx} =1$. In the above relations we have introduced 
 $\sigma({\bf x}) = \mid H({\bf x}) \mid^2$ represents the density of the condensate per unit volume around the point ${\bf x}$.

The boundary conditions for the equation (\ref{sa1}) are given by demanding that $A^{(1)}_t $ 
is equal zero at $u=0$ and for $u=1$. For the  equation (\ref{sa2})
we impose that the solution is regular on the event horizon and
$\ep_H~F_{xy}^{(1)}$ calculated at the boundary of the AdS spacetime as given by the difference between 
magnetic and critical magnetic field $B-B_{c2}$.

The formal solutions of the above relations can be cast in the form as
\ben \label{a1t}
A^{(1)}_t &=& - 2 \frac{r_+^2}{\tilde \alpha} \int_0^1 du'~\frac{\rho_0^2(u')}{u'^2}~A^{(0)}_t ~\int d{\bf x'} ~G_B(u,~u' \mid {\bf x} - {\bf x'})~\sigma({\bf x'}), \\ \label{a1i}
A^{(1)}_m &=& a_{m}({\bf x}) - \frac{r_+^2}{\tilde \alpha} ~\ep_m{}^{j}~\int_0^1 du'~\frac{\rho_0^2(u')}{u'^2}~\int d{\bf x'} ~G_s(u,~u' \mid {\bf x} - {\bf x'})~\p_j \sigma({\bf x'}), 
\een
where the Green functions connected with the time and spatial components of the Maxwell gauge potentials imply
\ben
D^{(t)}  G_B(u,~u' \mid {\bf x} - {\bf x'}) &=& - \delta(u -u')\delta( {\bf x} - {\bf x'}),\\
D^{(i)}  G_s(u,~u' \mid {\bf x} - {\bf x'}) &=& - \delta(u -u')\delta( {\bf x} - {\bf x'}),
\een
with the AdS boundary conditions provided by the following:
\ben
G_B(u,~u' \mid {\bf x} - {\bf x'}) \mid_{u=0} &=& G_B(u,~u' \mid {\bf x} - {\bf x'}) \mid_{u=1} = 0,\\
G_s(u,~u' \mid {\bf x} - {\bf x'}) \mid_{u=0} &=& G_s(u,~u' \mid {\bf x} - {\bf x'}) \mid_{u=1} = 0.
\een

\subsubsection{Current}
The AdS prescription enables us to find the $U(1)$-gauge fields current
\be
<J^{\beta}> = \frac{\delta S_{on-shell}}{\delta A_\beta} \mid_{u \rightarrow 0} = \bigg(F^{\beta u} + \frac{\alpha}{2} B^{\beta u}\bigg) \mid_{u \rightarrow 0}.
\ee
Using  equation (\ref{a1i}) for $A^{(1)}_j$ and $F_{j u}^{(1)} = - \p_u A_j^{(1)}$  we arrive at the folowing relation for the spatial components of the current in question
\be
<J_i> ^{DM}= \bigg( F_{i u} + \frac{\alpha}{2} B_{i u} \bigg) \mid_{u \rightarrow 0}.
\ee
In order to evaluate it let us consider the $u$-component of the first order expansions of the gauge fields in equation of motion (\ref{bb}). It implies
\be
\p_i \bigg[ \sqrt{-g} ~\bigg( B^{i u (1)} + \frac{\alpha}{2} F^{i u (1)} \bigg) \bigg]= 0.
\ee
Having in mind the line element describing our spacetime we obtain the relation for the covariant components of the gauge fields provided by
\be
B_{i u }^{(1)} + \frac{\alpha}{2} F_{i u}^{(1)}  = \frac{c_i~r_+}{u^2~f(u)},
\ee
where $c_i$ are constants bounded with $x$ or $y$ components, respectively. Next, extracting $B_{i u }^{(1)} $ from the last equation and inserting it into $<J_i> ^{DM}$ we get
\be
<J_i> ^{DM}= {\tilde \alpha}~F_{i u}^{ (1)} \mid_{u \rightarrow 0} + \beta_i (\alpha),
\ee
where we set
\be
\beta_i(\alpha) = \frac{\alpha~c_i}{2~r_+}.
\ee
By virtue of the above one finally arrives at the expression for the $U(1)$-gauge current in the theory with {\it dark matter} sector. Calculating $F_{i u}^{ (1)}$ using equation
(\ref{a1i}) implies
\be
<J_i>^{DM} = \ep_i{}^m ~\p_m \Theta({\bf x}) \mid_{u \rightarrow 0} + \beta_i(\alpha),
\label{cur}
\ee
where we have denoted
\be
\Theta({\bf x}) = r_+^2~\int_0^1 du'~\frac{\rho_0^2(u')}{u'^2}~\p_u \int d{\bf x'} ~G_s(u,~u' \mid {\bf x} - {\bf x'})~\sigma({\bf x'}) \mid_{u \rightarrow 0}.
\ee
One can see that the dependence on the {\it dark matter} sector (dependence on $\tilde \alpha$ ) cancels in the first term and the current in the presence of 
the {\it dark matter} sector is provided by
\be
 <J_a>^{DM} = <J_a> + ~\beta_a (\alpha).
 \ee
From the above relation it can be concluded that if we take an integration constants $c_i$ equal to zero, one receives that the {\it dark matter} sector current is the same as in the 'ordinary' case.

\section{Free energy}
\label{sec:fenergy}

In this section we shall find free energy of the considered system, i.e., Maxwell {\it dark matter} sector vortex configuration in $(2+1)$-dimensions. The free energy notion is important from the
point of view of the thermodynamical stability of the aforementioned configuration. It turns out that 
in holographic approach to the physical problems the free energy for the boundary theory
can be determined as the on-shell gravity action in the bulk. 

In our case the on-shell action bounded with the scalar field in question vanishes identically. It is caused by the fact that
the scalar field has a compact support and satisfies the boundary conditions of the form $\psi \sim c_1~u^2$, where $c_1$ is constant, at the AdS boundary.  By virtue of the above
one can restrict the attention to Maxwell  and {\it dark matter} $U(1)$-gauge fields.
\be
S_{on-shell} = - \int_{M} d^4x \sqrt{-g}~\bigg( \frac{1}{4} F_{\mu \nu}F^{\mu \nu} + \frac{\alpha}{4}F_{\mu \nu} B^{\mu \nu} \bigg).
\ee
Let us expand the above action in terms of $\ep_{H}$
\be
S_{on-shell} = S^{(0)} + \ep_H~S^{(1)} + \ep_H^2~S^{(2)} + \dots.
\label{onshell}
\ee
On the other hand, for the gauge fields one obtains
\be
F_{\mu \nu}^{(i)} = \p_\mu A_\nu^{(i)} - \p_\nu A_\mu^{(i)}, \qquad
B_{\mu \nu}^{(i)} = \p_\mu B_\nu^{(i)} - \p_\nu B_\mu^{(i)}, \qquad i = 0,~1,~2,\dots
\ee
The $S^{(0)} $ coefficient implies
\be
- S^{(0)}  = \int_{\cal M} d^4x~\sqrt{-g} \bigg( \frac{1}{4} F_{\mu \nu}^{(0)} F^{\mu \nu (0)}  + \frac{\alpha}{4}F_{\mu \nu}^{(0)} B^{\mu \nu (0)} \bigg) .
\label{so}
\ee
The first term on the right-hand side of (\ref{so}) accords to a trivial configuration without any scalar field, i.e., $\psi = 0$. However, the other terms in the expansion
described by the relation (\ref{onshell}) possess the information about states in which one has non-zero condensate.\\
Let us consider the $S^{(1)} $ coefficient which may be rewritten in the form as
\ben
- S^{(1)}  &=& \int_{\cal M} d^4x~\sqrt{-g} \bigg[ \frac{1}{2} F_{\mu \nu}^{(1)} F^{\mu \nu (0)}  + \frac{\alpha}{4} \bigg( F_{\mu \nu}^{(1)} B^{\mu \nu (0)}  +
F_{\mu \nu}^{(0)} B^{\mu \nu (1)} \bigg) \bigg] \\ \nonumber
&=& \int_{\p{\cM}} d \Sigma_{u} ~F^{u \nu (0)} A_{\nu}^{(1)} \mid_{u \rightarrow 0}
 + \frac{\alpha}{2} \bigg[ \int_{\p \cM} d \Sigma_{u} ~\bigg( B^{u \nu (0)} A_{\nu}^{(1)}  +
F^{u \nu (0)} B_{\nu}^{(1)} \bigg) \bigg]_{u \rightarrow 0},
\een
where we used the equations of motion for the zeroth order, i.e., $\na_{\mu} F^{\mu \nu(0)} = 0$ which in turn implies that $\na_{\mu} B^{\mu \nu(0)} = 0$. 
These terms vanish because of the fact that one has to take into account the boundary theory at some fixed value of the chemical potential $\mu$.

The $S^{(2)} $ coefficient is provided by the following relations:
\ben
- S^{(2)}  &= & \int_{\cal M} d^4x~\sqrt{-g} \bigg[ \frac{1}{2} F_{\mu \nu}^{(1)} F^{\mu \nu (0)}  + \frac{1}{4} F_{\mu \nu}^{(1)} F^{\mu \nu (1)} \\ \nonumber
&+& \frac{\alpha}{4} \bigg( F_{\mu \nu}^{(2)} B^{\mu \nu (0)}  + F_{\mu \nu}^{(1)} B^{\mu \nu (1)}  +
F_{\mu \nu}^{(0)} B^{\mu \nu (2)} \bigg) \bigg] \\ \nonumber
&=& \int_{\p \cM} d \Sigma_{u} ~\bigg( F^{u \nu (0)} A_{\nu}^{(2)}  + \frac{1}{2} F^{u \nu (1)} A_{\nu}^{(1)}  \bigg) \mid_{u \rightarrow 0} 
- \frac{3}{2}\int_{\cM} d^4x~\sqrt{-g}\frac{j^{\nu (1)}}{\tilde \alpha} A_\nu^{(1)} \\ \nonumber
&+& \int_{\p \cM} d \Sigma_{u} ~\frac{\alpha}{2} \bigg( B^{u \nu (0)} A_{\nu}^{(2)} + F^{u \nu (0)} B_{\nu}^{(2)} + F_{\mu \nu}^{(1)} B^{\mu \nu (1)} \bigg) \mid_{u \rightarrow 0} ,
\een
where one has used the equations of motion for the first order, i.e., ${\tilde \alpha}\na_{\mu} F^{\mu \nu (1)} = j^{\nu (1)}$.
The 'orthogonality condition' \cite{mae10}
\be
\int_{\cM}d^4x~\sqrt{-g}~A_\mu^{(1)}~j^{\mu (1)} = 0,
\ee
can be implemented and the second term in the third line vanishes.
Consequently we arrive at the following expression:
\be
- S_{on-shell} = \int_{\p \cM} d \Sigma_{u} ~\bigg[ \frac{\alpha}{2} 
\bigg( F^{u \nu (0)} B_{\nu} + B^{u \nu (0)} A_{\nu} \bigg) + \frac{\ep^2}{2} A_{\nu}^{(1)} \bigg( F^{u \nu (1)} + \alpha B^{u \nu (1)} \bigg)
\bigg] + \cO(\ep_H^3).
\label{ons}
\ee
The same arguments as applied in $S^{(1)}$ analysis, lead us to the conclusion that the first term on the right-hand side of equation (\ref{ons}) vanishes. 
In the next step we use the fact that
$A_t^{(1)}({\bf x},~0) = 0$. Consequently, by virtue of the above the on-shell action is provided by the relation of the form
\be
- S_{on-shell} = \frac{\ep^2}{2} \int d^3 x~\delta^{ab} \bigg( F_{u a}^{ (1)} + \alpha B_{u a} ^{(1)} \bigg)~A_{b}^{(1)}  \mid_{u \rightarrow 0} +  \cO(\ep_H^3).
\ee
Having in mind the relation for $<J_a>^{DM}$ one has that
\be
- S_{on-shell} = \frac{\ep^2}{2} \int d^3 x~\delta^{ab} ~<J_a>^{DM}~A_b^{(1)} \mid_{u \rightarrow 0} + \cO(\ep_H^3).
\ee
Putting together all those results, we obtain that the expression describing free energy for the considered theory with {\it dark matter} sector in $(2+1)$-dimensions. It yields
\ben
F = - \frac{\ep^2}{2} \int_{\cR^2} d^2 {\bf x}~\delta^{ab} ~<J_a>^{DM}~A_b^{(1)} \mid_{u \rightarrow 0}  = 
&-& \frac{\ep^2}{2}~B_{c_2} ~\int_{\cR^2} d^2 {\bf x} {~\Theta(\bf x)} \\ \nonumber
 &-& \frac{\ep^2}{2} \int_{\cR^2} d^2{\bf x}~\beta_i(\alpha)~A^{i (1)},
\een
where $B_{c_2}$ stands for 'the upper magnetic' $U(1)$-gauge Maxwell field.\\
Let us suppose that we take into account some bounded region $V$ of the two-dimensional hypersurface $\cR^2$ in question. The free energy density 
may be written as
\be
{\cal F} = \frac{F}{V} = - \frac{\ep^2}{2}~B_{c_2}{~<\Theta(\bf x)>}  - \frac{\ep^2}{2} ~<\beta_i(\alpha)~A^{i (1)}>,
\ee
where $<\Theta(\bf x)>$ and $<\beta_i(\alpha)~A^{i (1)}>$ denote the averages of $\Theta({\bf x})$ and $\beta_i(\alpha)~A^{i (1)}$ over the two-dimensional hypersurface $\cR^2$. 

It is worth pointing out that the free energy density consists of two terms, first one is connected with the ordinary Maxwell field while the other one is bounded with {\it dark matter} sector.
Because of the fact that $\cal F$ is negative one can conclude that
the considered system is more stable over the trivial configuration where there is no charge condensate, 
i.e., the value of the scalar field  $\psi$ is equal to zero.
Moreover, the free energy is smaller for the {\it dark matter} sector, which means that the vortices create more easily in the presence of {\it dark matter} than without it. Perhaps this could 
be the answer why the {\it cosmic web } of tendrils (resembling vortices in the large cosmological scales) form first and then an ordinary matter condenses on them to form galaxies and clusters of galaxies.

\section{Superconducting coherence length}
\label{sec:coherence}
We turn now to the analysis of the superconducting length and the influence of the {\it dark matter} sector on it. 
In our analysis we shall consider the following components of the Maxwell and scalar fields:
\be
A_\mu = (\phi(u),~0,~0,~A_y), \qquad \psi = \psi(u).
\ee
By virtue of the above choice we can straightforwardly verify that the equations of motion imply
\ben 
\label{eqm1a}
\p^2_u \phi(u) &-& \frac{2~r_+~\psi^2~\phi}{\tilde \alpha~u^4~f(u)} = 0,\\ 
\label{eqm2a}
\p_u \bigg( f(u)~\p_u \psi \bigg) &-& \frac{m^2~r_+^2}{u^4}\psi + \frac{\phi^2~r_+^2}{f(u)~u^4} \psi - \frac{1}{u^4}~A_y^2~\psi = 0.
\een

It happens that the superconducting coherence length $\xi$ can be connected with the correlation length of the order parameter in the momentum space \cite{mae08}.
$\chi$ emerges as the pole of the static correlation function of the order parameter in the Fourier space. Namely, one can write that
\be
<\cO(k)~\cO(-k)> \sim \frac{1}{\mid k \mid^2 + \frac{1}{\xi^2}}.
\ee
As in \cite{mae08} the pole of the static correlation function of a dual field operator can be found by solving
the eigenvalue problem for the static perturbation subject to the condition that the wave number $k$ of the corresponding bulk field is given by
$1/\xi^2 = - {k_\ast}^2$, where $k_*$ is a wave number allowed as eigenvalues \cite{mae08}.
The aforementioned condition gives us the pole of the static correlation function.
Because of the fact that near the critical point, when $T \rightarrow T_c$, the coherence length is divergent, one ought to solve the eigenvalue problem
being subject  to the condition
\be
\lim_{\ep_T \rightarrow 0} (-k^2) = -{k_\ast}^2 = \frac{1}{\xi^2} = 0,
\ee
where $\ep_T = (T_c-T)/T_c$ with the auxiliary demand that $\mid \ep_T \mid \ll 1.$
In order to solve the equations of motion let us perform
the perturbative analysis in the fluctuations of the fields under considerations given by
\ben \label{field}
\phi &=& \varphi(u) + \ep_T~A^{(1)}_t (u) + \cO(\ep_T^2),\\
\psi &=& \sqrt{\ep_T}~\psi_1(u) + \ep_T^{\frac{3}{2}} ~\psi_2(u)+ \cO(\ep_T^2).
\een
In the leading order of the $\ep_T$ parameter, the underlying  
equations of motion can be cast in the forms as stated below.
Namely, in the first order of $\ep_T$-order we get
\be
\p^2_u A^{(1)}_t (u) - \frac{2~r_+^2}{{\tilde \alpha}~u^4~f(u)} \psi_1^2(u) \varphi(u) = 0.
\label{e1}
\ee
On the other hand, at $\sqrt{\ep_T}$-order they imply
\be
\p_u \bigg( f(u)~\p_u \psi_1(u) \bigg) - \frac{m^2~r_+^2}{u^4} \psi_1(u) + \frac{r_+^2~\varphi^2(u) ~\psi_1(u)}{f(u)~u^4} - \frac{1}{u^4}~A_y^2~\psi_1(u) = 0,
\label{e11}
\ee
and at $\ep_T^{\frac{3}{2}}$-order, one arrives at the relation as follows:
\be
\p_u \bigg( f(u)~\p_u \psi_2(u) \bigg) - \frac{m^2~r_+^2}{u^4} \psi_2(u) + \frac{2~r_+^2~\varphi(u) ~\psi_1(u)~A^{(1)}_t (u)}{f(u)~u^4} = 0.
\label{e111}
\ee

In order to determine the coherence length, we shall elaborate fluctuations around the background fields $A_\mu(u)$ and $\psi(u)$.
It turned out that static perturbations will be adequate for our purpose. Thus, we pay attention to the fluctuations with only one spatial direction along $x$-direction.
The linear perturbations of gauge fields and scalar one, will yield
\ben \label{a1}
\delta A_\mu &=& \bigg[ a_t(u,~k) dt + a_x(u,~k) dx + a_y(u,~k) dy \bigg]~ e^{ikx},\\ \label{a2}
\delta \psi &=& \bigg[ b(u,~k) + i~{\bar b}(u,~k) \bigg]~e^{ikx}.
\een
The boundary conditions for the gauge and scalar fields should be imposed. Namely, let us suppose that at the black brane horizon $(u=1)$ and near of the 
AdS spacetime boundary $(u = 0)$, the asymptotic behavior of the fields in question imply
\ben
a_t(u,~k) \mid_{u \rightarrow 1} &=& 0,\\
b(u,~k) \mid_{u \rightarrow 1} &=& regular,\\
a_t(u,~k) \mid_{u \rightarrow 0} &=& const~u,\\
b(u,~k) \mid_{u \rightarrow 0} &=& const~u^2.
\een
Intersecting (\ref{a1})-(\ref{a2}) into the underlying equations of motion provides the following:
\ben
k^2~a_t = f(u)~u^2~\p^2_u a_t - \frac{2~\psi^2~r_+^2a_t}{{\tilde \alpha}~u^2} &-& \frac{4~\psi~b~r_+}{{\tilde \alpha}~u^2} \phi = 0,\\ \label{lon}
k^2~a_y = \p_u \bigg[ u^2~f(u)~\p_u a_y \bigg] - \frac{2~\psi^2~r_+^2~a_y}{{\tilde \alpha}~u^2} &-& \frac{4~\psi~b~A_y~r_+^2}{{\tilde \alpha}~u^2} = 0,\\
k^2~b = u^2~\p_u \bigg[ f(u) \p_u b \bigg] &+& \frac{r_+^2 ~b~\phi^2}{f(u)~u^2} - A_y^2~b -\frac{m^2~r_+^2~b}{u^2}\\ \nonumber
&+& \frac{2~a_t~\phi~\psi~r_+^2}{f(u)~u^2} - 2~a_t~A_y~\psi = 0,\\
\p_u \bigg[ u^2~f(u)~\p_u a_x \bigg] &-& \frac{2~r_+^2}{{\tilde \alpha}~u^2} \psi^2~a_x = 0.
\een
The Wick rotation of the wave vector $k$, ~
$k \rightarrow ik$ and substitution the relation (\ref{field}) reveal the following:
\ben \label{at1}
- k^2~a_t = f(u)~u^2~\p^2_u a_t - \frac{2~ \ep_T~~\psi_1^2~r_+^2a_t}{{\tilde \alpha}~u^2} &-& \frac{4~\sqrt{\ep_T}~\psi_1~b~r_+~\varphi}{{\tilde \alpha}~u^2} \phi = 0,\\ \label{bt1}
- k^2~b = \bigg( D_b + \frac{2~\ep_T~r_+^2~\varphi~A^{(1)}_t (u) }{f(u)~u^2} \bigg) b &+& \frac{2~\sqrt{\ep_T}~a_t~\varphi~\psi_1~r_+^2}{f(u)~u^2} \\ \nonumber
&-& 2~\sqrt{\ep_T}~a_y~A_y~\psi_1 = 0,
\een
where we have define the differential operator $D_b$ by the relation 
\be
D_b = u^2~\p_u\bigg( f(u)~\p_u \bigg) + \frac{r_+^2~\varphi^2}{f(u)~u^2} - A_y^2 - \frac{m^2~r_+^2}{u^2}.
\ee
The zeroth order solutions of the set of equations (\ref{at1})-(\ref{bt1}), which are consistent with the asymptotical behavior near the black brane event horizon and AdS spacetime boundary,
have the forms given by the relations
\be
b^{(0)} = \psi_1, \qquad a^{(0)}_t = 0, \qquad \qquad a^{(0)}_y= 0.
\ee
The zeroth order solutions are in accord with the fact that at this order they are only functions of $u$-coordinate.
Having in mind that perturbations described by the equations (\ref{a1})-(\ref{a2}) have the spatial dependence on $x$-direction,
we conclude that the first non-trivial corrections emerges at $\sqrt{\ep_T}$-order . Thus, we suppose that
\be
a_t = \sqrt{\ep_T}~{\tilde a}_t, \qquad a_y= \sqrt{\ep_T}~{\tilde a}_y,
\ee
and substitute the above expressions into the relations for $a_t,~a_y,~b$. Consequently, one obtains
\ben \label{x1}
- k^2~{\tilde a}_t &=& f(u)~u^2~\p^2_u {\tilde a}_t - \frac{4~\psi_1~b~r_+^2~\varphi}{{\tilde \alpha}~u^2} - 2~\ep_T~\frac{\psi_1^2~r_+^2}{{\tilde \alpha}~u^2}~{\tilde a}_t = 0,\\ \label{x2}
- k^2~b &=& D_b b+ \ep_T \bigg(
 \frac{2~r_+^2~\varphi~A^{(1)}_t (u) ~b}{f(u)~u^2} + \frac{2~{\tilde a}_t~\varphi~\psi_1~r_+^2}{f(u)~u^2}  - 2~{\tilde a}_y~A_y~\psi_1 \bigg) = 0,\\
 - k^2~{\tilde a}_y &=& \p_u \bigg( u^2~f(u)~\p_u {\tilde a}_y \bigg) - \frac{4~\psi_1^2~A_y~r_+^2}{{\tilde \alpha}~u^2} + \cO(\ep_T^{n \ge 1/2}).
 \een
 Moreover we write down the set of expansions for the coefficients of the static perturbations
 \ben
 b &=& \psi_1 + \ep_T~b^{(1)} + \cO(\ep_T^2),\\
 {\tilde a}_t &=& {\tilde a}^{(0)}_t  + \cO(\ep_T),\\
 {\tilde a}_y &=& {\tilde a}^{(0)}_y  + \cO(\ep_T),\\
{k}_\ast^2 &=& \ep_T~{k}_1^2 + \cO(\ep_T^2).
\een
By virtue of the equations (\ref{x1})-(\ref{x2}) as well as (\ref{e1})-(\ref{e11}), taking into account the limit $\ep_T \rightarrow 0$, which means that
$k^2 \rightarrow k_\ast^2$,  we can readily verify that
\ben
\p^2_u {\tilde a}^{(0)}_t (u,~{k}_\ast) &=& \frac{4~\psi^2~r_+^2~\varphi}{{\tilde \alpha}~f(u)~u^4} = 2~\p_u^2 A^{(1)}_t (u) ,\\
- {k}_1^2~\psi_1 &=& D_b b^{(1)}(u,~{k}_\ast) \\ \nonumber
&+& \bigg(
\frac{2~r_+^2~\psi_1~\varphi~A^{(1)}_t (u) }{f(u)~u^2} + \frac{2~{\tilde a}^{(0)}_t  ~\varphi~\psi_1~r_+^2}{f(u)~u^2} - 2~{\tilde a}^{(0)}_y~A_y~\psi_1 \bigg) = 0.
\een
In order to calculate the superconducting coherence length we define the scalar product defined by
\be
<\psi_1 \mid \psi_1> = \int_0^1 du ~\frac{\psi_1^\ast ~\psi_1}{u^2}.
\ee
The form of the scalar product reveals the fact that the operator $D_b$ is the Hermitian one. 
In the next step, we multiply the equation for ${\tilde k}_1$ by the bra $<\psi_1 \mid$. On this account it is customary to write
\ben
- {k}_1^2~< \psi_1 \mid \psi_1 > &=& <\psi_1 \mid D_b b^{(1)}> + < \psi_1 \mid \frac{2~r_+^2~\psi_1~\varphi~A^{(1)}_t (u) }{f(u)~u^2} >  \\ \nonumber
 &-& \frac{\tilde \alpha}{2}
\int_0^1 du~\bigg(\frac{d{\tilde a}^{(0)}_t }{du} \bigg)^2 
- 2~A_y~\int_0^1 \frac{du}{u^2}~\psi_1^2~{\tilde a}^{(0)}_y .
\een
The first term on the right-hand side can be found considering $\ep_T^{3/2}$-order behavior of the equations of motion 
for the underlying theory. Namely the relation (\ref{e111}) enables us to write 
\be
D_b \psi_2 = - \frac{2~r_+^2~\psi_1~\varphi~A^{(1)}_t (u) }{f(u)~u^2},
\ee
which in turn leads us to the conclusion that $<\psi_1 \mid D_b \psi_2> = 0$.\\
Finally, coming back to the original vector wave, by performing the Wick rotation as well as 
taking into account the limit when $\ep_T \rightarrow 0$, we receive the searched for expression
\be
 - {k}_1^2 = \frac{\cal P}{\cal M}.
 \ee
 For the brevity of the notation we have defined the above quantities ${\cal P}$ and $\cal M$ as
 \ben
 \cal P &=& \frac{\tilde \alpha}{2}~\int_0^1du~\bigg( \frac{d {\tilde a}^{(0)}_t }{d u} \bigg)^2 + 2~A_y~\int_0^1 du~\psi_1^2~{\tilde a}^{(0)}_y,\\
\cal M &=& \int_0^1 du ~\frac{\psi_1^\ast ~\psi_1}{u^2}.
\een
On this account the superconducting coherence length in the theory with {\it dark matter} sector yields
\be
\xi = \sqrt{\frac{\cal M}{\cal P}}~\bigg( 1 - \frac{T}{T_c} \bigg)^{\frac{1}{2}} \sim {\tilde \alpha}^{- \frac{1}{2}}~\bigg( 1 - \frac{T}{T_c} \bigg)^{\frac{1}{2}} .
\ee
The coherence length depends on the $\alpha$-coupling constant (which in turn describes the dependence on the {\it dark matter} model). 
The main conclusion of these calculations is that the smaller $\alpha$ we consider, the 
greater value of the superconducting length we obtain.

\section{London equation with {\it dark matter} sector}
\label{sec:london eq}
This section is devoted to calculations of the magnetic penetration depth and the number density 
of the superfluid  for Maxwell {\it dark sector} 
vortices in $(2+1)$-dimensions in the presence of a homogeneous external magnetic field. The magnetic field is chosen as perpendicular to the two-dimensional hypersurface. Let us introduce 
the ansatz of the form as
\be
\delta A_y(u,~x) = a_y(u)~x,
\label{ay}
\ee
which is equivalent to the perturbation of the gauge field with the wave vector $k=0$.\\
As was mentioned in section 2, $A_y$ can be connected with the {\it dark matter} magnetic field or Maxwell magnetic field. For generality of our considerations we set in what follows $A_y$.
Inserting it into equation (\ref{lon}) one readily writes down
\be
\p_u \bigg( u^2~f(u)~\p_u a_y \bigg) - \frac{2~r_+^2~\psi^2~a_y}{{\tilde \alpha}~u^2} - \frac{4~\psi~b~A_y~r_+^2}{{\tilde \alpha}~u^2}  = 0.
\ee
As in the previous section we assume the following perturbative expansion:
\ben
\psi &=& \sqrt{\ep_T}~\psi_1(u) + \ep_T^{3/2}~\psi_2(u) + \cO(\ep_T^{5/2}),\\
a_y &=& a^{(0)}_y + \ep_T~a^{(1)}_y + \cO(\ep_T^{3}),\\
b &=& \psi_1(u) + \sqrt{\ep_T}~b^{(1)} + \cO(\ep_T).
\een
Up to $\ep_T^0$ and to $\ep_T^1$ orders the equations are given respectively by
\ben \label{l1}
\p_u \bigg( u^2~f(u)~\p_u a^{(0)}_y \bigg)  &=& 0,\\ \label{l2}
\p_u \bigg( u^2~f(u)~\p_u a^{(1)}_y \bigg) - \frac{2~r_+^2}{{\tilde \alpha}~u^2}\psi_1^2(u)~a^{(0)}_y &-& \frac{4~b^{(1)}~\psi_1(u)~r_+^2}{{\tilde \alpha}~u^2} A_y = 0.
\een
From the equation (\ref{l1}) it can be seen that $a^{(0)}_y $ is constant and $a^{(1)}_y $ solution can be formally written  as
\ben
a^{(0)}_y  &=& C_0 = const,\\
a^{(1)}_y &=& C_1 - 2~C_0~\int_0^u \frac{du''}{1 - u''^3}~\int_{u''}^1 \frac{du'}{\tilde \alpha~u'^2}~\psi_1(u')~\bigg( \psi_1(u') + \frac{2~b^{(1)}(u')~A_y}{C_0} \bigg) + \cO(\ep_T^2).
\een 
To proceed further, we set $C_1=0$ and $C_0 = A_y$. On this account the relation (\ref{ay}) for $\delta A_y(u,~x)$ implies
\be
\delta A_y (u,~x) \simeq \delta A^{(0)}_y (x)~\bigg( 1 - 2~\ep_T~u~\int_{u''}^1 \frac{du'}{{\tilde \alpha}~u'^2} \psi_1(u')~\bigg(
\psi_1(u') + 2~b^{(1)}(u') \bigg)\bigg)+ \cO(\ep_T^2),
\ee
where $\delta A^{(0)}_y = \lim_{u \rightarrow 0} \delta A_y(U,~x) = A_y$.\\
On the other hand, from the asymptotic behavior of $\delta A_y$ near the AdS spacetime boundary
\be
\delta A_y = \delta A_y^{(0)} + \frac{\delta A_y^{(1)}}{r} + \dots,
\ee
and taking into account the definition $r_+ = 4 \pi~{T_c }/3$, we can read off the form of $<J_y(x)>$ when
$T \rightarrow T_c$. Namely, it is given by the relation
\be
<J_y(x)>_{u \rightarrow 0} = - \frac{8\pi~\ep_T}{3}~T_c~\delta A^{(0)}_y (x)~\int_0^1\frac{du}{{\tilde \alpha}~u^2}~{\cal K}(u,~k) + \cO(\ep_T^2),
\label{jcur}
\ee
where we have denoted by ${\cal K}(u,~k) $ the relation provided by
\be
{\cal K}(u,~k) = \psi_1^2(u) + 2~\psi_1(u)~b^{(1)}.
\ee
Let us find the behavior of the integral (\ref{jcur}) near the AdS spacetime boundary. Near the boundary (for $m^2 = -2$) the scalar field $\psi_1$ is proportional to the condensation operator
$<\cO>$. Because of  the fact that we reach almost the critical temperature, the condensation operator plays the role of the order parameter for the boundary theory.
In general, the considered scalar field solution can be expressed as a function of the $u$-coordinate \cite{mae08} in the form as
\be
\psi(u) = \psi^m~f(u),
\ee
where $f(u)$ is the solution of equation satisfying the limit $\lim_{u \rightarrow 0} f(u) = u^{\Delta_m}$, 
 and regularity condition at the black hole event horizon. By virtue of this, we set
\be
\psi_1 \sim <\cO>~u^2.
\ee
For the upper limit in the considered integral, i.e., for $u=1$, $\psi_1$ is perfectly regular. All these facts suggest that the integral in question is well defined and has the finite value.

As was revealed in \cite{mae10} 
the form of $<J_y>$ closely resembles the expression from the Ginzburg-Landau theory, so-called London equation,
in which the order parameter $\psi$ is coupled to the $U(1)$-gauge field $A_a$ and the current
$J_a$ is provided by
\be
J_a = - \frac{e^2}{m} \psi^2~A_a = - e~n_s~A_a,
\ee
where $e$ and $m$ are effective charge and mass of the order parameter. $n_s$ is connected with the superfluid number density.
In the expression for the average spatial current in y-direction, $\delta A_y^{(0)}$ plays the role of an external source. On the contrary, in London equation $A_a$
is built from the spatial average of the microscopic field as well as an external field. In the considered attitude we have no dynamical photon (the current does not produce its own magnetic field).
Summing it all up, it means that the external $U(1)$-gauge field $\delta A_y^{(0)}$ equals to the macroscopic gauge field in the AdS/CFT attitude. So the
comparison of $J_a$ and $<J_y>$ reveals that,
near the boundary of the AdS spacetime we have that the number density of the superfluid particles is given by
\be
n_s ^{DM}= \frac{8 \pi ~T_c}{3~{\tilde \alpha}}~<\cO>^2~F(u,~k) = \frac{n_s}{\tilde \alpha},
\label{ns}
\ee
where $F(u,~k)$ stands for the integral
\be
F(u,~k) = \int_0^1\frac{du}{u^2}~{\cal K}(u,~k).
\ee
On the other hand, the magnetic penetration depth implies
\be
\lambda^{DM} = \sqrt{{\tilde \alpha}}~\lambda.
\label{ds}
\ee
Inspection of the equations (\ref{ns}) and (\ref{ds}) reveals that the bigger $\alpha$-coupling constant one takes 
the greater $n_s ^{DM}$ we get. On the contrary, the smaller $\alpha$  the greater magnetic penetration we have.\\
It is interesting to note that {\it dark matter} affects both the penetration depth and coherence length
in the same way leaving their ratio intact $\lambda^{DM}/\xi^{DM}=\lambda/\xi$. Noting that this ratio
decides if the superconductor is of first or second type, the conclusion is that, at least in the probe
limit, the dark matter does not affect classification of the holographic superconductors.

\section{Conclusions}
\label{sec:conclusions}
In the paper we have considered AdS/CFT - gauge/gravity, correspondence in order to study linear fluctuations of scalar field solution 
in the s-wave holographic superconductor under the assumption of the probe limit, i.e., the fluctuations do not backreact on the gravitational field. We have analyzed the theory
in which $U(1)$-gauge Maxwell field is coupled with the other gauge field which mimics the presence of the {\it dark matter} sector. 

The main aim of the work was to answer the question how
{\it dark matter}  sector ( $\alpha$ coupling constant of these aforementioned gauge fields) will modify characteristics of vortex solutions, like the coherence length,  superfluid density and magnetic penetration depth.
It happened that both, the coherence length and the penetration depth were affected by the presence of the {\it dark matter} 
sector in the same way.  The smaller value of $\alpha$ is taken into account the greater value of the  $\xi$ and $\lambda$ is obtained.
On the other hand, the smaller $\alpha$ we have, the greater superfluid density we get.

We have also found that the free energy for the vortex configuration turns out to be negative and  for the configuration 
with {\it dark matter} sector lesser than in the 'ordinary' Maxwell case.
This fact enables us to conclude that the vortex configurations are stable over the trivial configuration where there is no charge scalar condensate. Secondly, 
because of the fact that in the presence of {\it dark matter} the vortices build less free energy configuration 
than ordinary matter ones (e.g., Maxwell or Chern-Simons), it can give rise to the answer why in the Early Universe
first  the web of {\it dark matter} formed and next on these tendrils the  ordinary matter condensed forming cluster of galaxies.





\acknowledgments
MR was partially supported by the grant of the National Science Center \\
$DEC-2014/15/B/ST2/00089$
and KIW by the grant DEC-2014/13/B/ST3/04451.




\begin{thebibliography}{99}

%
\def\cmp#1#2#3#4{\emph{#4}, \emph{ Commun. Math. Phys.} {\bf #1} (#3) #2}
\def\lmp#1#2#3#4{\emph{#4}, \emph{ Lett. Math. Phys.} {\bf #1} (#3) #2}
\def\hpa#1#2#3#4{\emph{#4}, \emph{ Hell. Phys. Acta} {\bf #1} (#3) #2}
\def\grg#1#2#3#4{\emph{#4}, \emph{ Gen. Rel. Grav.} {\bf #1} (#3) #2}
\def\pr#1#2#3#4{\emph{#4}, \emph{ Phys. Rev.} {\bf #1} (#3) #2}
\def\prl#1#2#3#4{\emph{#4}, \emph{ Phys. Rev. Lett.} {\bf #1} (#3) #2}
\def\prd#1#2#3#4{\emph{#4}, \emph{ Phys. Rev. D} {\bf #1} (#3) #2}
\def\pl#1#2#3#4{\emph{#4}, \emph{ Phys. Lett.} {\bf #1} (#3) #2}
\def\pla#1#2#3#4{\emph{#4}, \emph{ Phys. Lett. A} {\bf #1} (#3) #2}
\def\plb#1#2#3#4{\emph{#4}, \emph{ Phys. Lett. B} {\bf #1} (#3) #2}
\def\prep#1#2#3#4{\emph{#4}, \emph{ Phys. Reports} {\bf #1} (#3) #2}
\def\phys#1#2#3#4{\emph{#4}, \emph{ Physica} {\bf #1} (#3) #2}
\def\jcp#1#2#3#4{\emph{#4}, \emph{ J. Comput. Phys.} {\bf #1} (#3) #2}
\def\jmp#1#2#3#4{\emph{#4}, \emph{ J. Math. Phys.} {\bf #1} (#3) #2}
\def\jpm#1#2#3#4{\emph{#4}, \emph{ J. Phys. A: Math. Gen.} {\bf #1} (#3) #2}
\def\cpr#1#2#3#4{\emph{#4}, \emph{ Computer Phys. Rept.} {\bf #1} (#3) #2}
\def\cqg#1#2#3#4{\emph{#4}, \emph{ Class. Quant. Grav.} {\bf #1} (#3) #2}
\def\cma#1#2#3#4{\emph{#4}, \emph{ Computers Math. Applic.} {\bf #1} (#3) #2}
\def\mc#1#2#3#4{\emph{#4}, \emph{ Math. Compt.} {\bf #1} (#3) #2}
\def\apj#1#2#3#4{\emph{#4}, \emph{ Astrophys. J.} {\bf #1} (#3) #2}
\def\apjs#1#2#3#4{\emph{#4}, \emph{ Astrophys. J. Suppl.} {\bf #1} (#3) #2}
\def\acta#1#2#3#4{\emph{#4}, \emph{ Acta Astronomica} {\bf #1} (#3) #2}
\def\apl#1#2#3#4{\emph{#4}, \emph{ Ann. Physik. (Leipzig)} {\bf #1} (#3) #2}
\def\amjp#1#2#3#4{\emph{#4}, \emph{Am. J. Phys.} {\bf #1} (#3) #2}
\def\anp#1#2#3#4{\emph{#4}, \emph{ Ann. Phys.} {\bf #1} (#3) #2}
\def\sa#1#2#3#4{\emph{#4}, \emph{ Sov. Astro.} {\bf #1} (#3) #2}
\def\sia#1#2#3#4{\emph{#4}, \emph{ SIAM J. Sci. Statist. Comput.} {\bf #1} (#3) #2}
\def\aa#1#2#3#4{\emph{#4}, \emph{ Astron. Astrophys.} {\bf #1} (#3) #2}
\def\mnras#1#2#3#4{\emph{#4}, \emph{ Mon. Not. R. Astr. Soc.} {\bf #1} (#3) #2}
\def\npb#1#2#3#4{\emph{#4}, \emph{ Nucl. Phys. B} {\bf #1} (#3) #2}
\def\prsla#1#2#3#4{\emph{#4}, \emph{ Proc. R. Soc. London, Ser. A} {\bf #1} (#3) #2}
\def\jhep#1#2#3#4{\emph{#4}, \emph{ JHEP} {\bf #1} (#2) #3}
\def\nuc#1#2#3#4{\emph{#4}, \emph{ Nuovo Cimento B } {\bf #1} (#3) #2}
\def\ijmp#1#2#3#4{\emph{#4}, \emph{ Int. J. Mod. Phys. D} {\bf #1} (#3) #2}
\def\atmp#1#2#3#4{\emph{#4}, \emph{ Adv. Theor. Math. Phys.} {\bf #1} (#3) #2}
\def\ptps#1#2#3#4{\emph{#4}, \emph{ Prog. Theor. Phys. Suppl.} {\bf #1} (#3) #2}
\def\ptp#1#2#3#4{\emph{#4}, \emph{ Prog. Theor. Phys.} {\bf #1} (#3) #2}
\def\lmp#1#2#3#4{\emph{#4}, \emph{ Lett. Math. Phys.} {\bf #1} (#3) #2}
\def\cpam#1#2#3#4{\emph{#4}, \emph{ Comm. Pure Appl. Math.}  {\bf #1} (#3) #2}
\def\adv#1#2#3#4{\emph{#4}, \emph{ Adv. Phys.}  {\bf #1} (#3) #2}
\def\zh#1#2#3#4{\emph{#4}, \emph{ Zh. Eksp. Teor. Fiz.}  {\bf #1} (#3) #2}

\def\jams#1#2#3#4{\emph{#4}, \emph{ J. Austral. Math. Soc. B} {\bf #1} (#3) #2}
\def\appa#1#2#3#4{\emph{#4}, \emph{ Acta Phys. Polonica A} {\bf #1}, (#3) #2}
\def\nat#1#2#3#4{\emph{#4}, \emph{Nature} {\bf #1}, (#3) #2}
\def\science#1#2#3#4{\emph{#4}, \emph{Science} {\bf #1}, (#3) #2}
\def\arcmp#1#2#3#4{\emph{#4}, \emph{Annual Rev. of Cond. Matter Physics} {\bf #1}, (#3) #2}
%
\def\hepph#1#2{{ hep-ph }{#1} (#2)}
\def\hepth#1#2{{ hep-th }{#1} (#2)}
\def\grqc#1#2{{ gr-qc }{#1} (#2)}
\def\ibid#1#2#3#4{\emph{#4}, {\it ibid.} {\bf #1} (#3) #2}
\def\conphy#1#2#3#4{\emph{#4}, \emph{Contemporary Physics} {\bf #1}, (#3) #2}
%
\bibitem{mal}
J.M.Maldacena, \atmp{2}{231}{1998}{The large-N limit of superconformal field theories and supergravity}.

\bibitem{wit98}
E.Witten, \atmp{2}{253}{1998}{Anti-de-Sitter space and holography}.
\bibitem{gub98}
S.S.Gubser, I.R.Klebanov and A.M.Polyakov, \plb{428}{105}{1998}{Gauge theory correlators from noncritical string theory}.
\bibitem{rev1} 
J.P.Gauntlett, J.Sonner and T.Wiseman
\prl{103}{151601}{2009}{Holographic Superconductivity in M Theory}.

\bibitem{sac12} 
S.Sachdev, \arcmp{3}{9}{2012}{What can gauge-gravity duality teach us about condensed matter physics?}.
\bibitem{gre13} 
A.G. Green, \conphy{54}{33}{2013}{ An Introduction to Gauge Gravity Duality and Its Application in Condensed Matter}.

\bibitem{har08}
S.A.Hartnoll, C.P.Herzog and G.T.Horowitz, \prl{101}{031601}{2008}{Building a holographic superconductor}.


\bibitem{che10}
J.W.Chen, Y.J.Kao, D.Maity, W.Y.Wen and C.P.Yeh, \prd{81}{106008}{2010}{Towards a holographic model of D-wave superconductors}.
\bibitem{ben10}
F.Benini, C.P.Herzog, R.Rahman and A.Yarom, \jhep{11}{2010}{137}{Gauge gravity duality for d-wave superconductors: prospects and chalanges}.
\bibitem{rog14a}
M.Rogatko and K.I.Wysoki\'nski, \appa{126}{A9}{2014} {Remarks on the Hall conductivity in chiral superconductors: weak vs. strong coupling approach}.
\bibitem{zen10}
H.B.Zeng, Z.Y.Fan and H.S.Zong, \prd{82}{126008}{2010}{d-wave holographic superconductor vortex lattice and non-Abelian holographic superconductor droplet}.


\bibitem{gub08}
S.S.Gubser and S.S.Pufu, \jhep{11}{2008}{033}{The gravity dual of a p-wave superconductor}.
\bibitem{hor10}
G.T.Horowitz and B.Way, \jhep{11}{2010}{011}{Complete phase diagrams for a holographic superconductor}.
\bibitem{bas10}
P.Basu, J.He, A.Mukharjee and H.H.Shieh, \plb{689}{45}{2010}{Hard-gapped holographic superconductors}.
\bibitem{apr11}                  
F.Aprile, D.Rodriguez-Gomez and J.G.Russo, \jhep{01}{2011}{056}{p-wave holographic superconductors and five-dimensional gauged supergravity}.
\bibitem{gan12}
S.Gangopadhyay and D.Roychowdhury, \jhep{08}{2012}{104}{Analytic study of properties of holographic p-wave superconductors}.
\bibitem{amm10}
M.Ammon, J.Erdmenger, V.Grass and P.Kerner, \plb{686}{192}{2010}{On holographic p-wave superfluids with back-reaction}.
\bibitem{liu15}
S.Liu, Y.Q.Wang, {\it Holographic model of hybrid and coexisting s-wave and p-wave 
Josephson junction}, \hepth{1504.0691}{2015}.




\bibitem{cai11b}
R.G.Cai, L.Li, H.Q.Zhang and Y.L.Zhang, \prd{84}{126008}{2011}{Magnetic field effect on the phase transition in AdS soliton spacetime}.
\bibitem{cai11c}
R.G.Cai, H.F.Li, H.Q.Zhang, \prd{83}{126007}{2011}{Analytical studies on holographic insulator/superconductor phase transitions}.
\bibitem{akh11}
A.Akhavan and M.Alishahiha, \prd{83}{086003}{2011}{p-wave holographic insulator/superconductor phase transition}.
\bibitem{amo14}
A.Amoretti, A.Braggio, N.Maggiore, N.Magnoli and D.Musso, \jhep{01}{2014}{054}{Coexistence of two vector order parameters: a holographic model for ferromagnetic superconductivity}.
\bibitem{sio10}
G.Siopsis and J.Therrien, \jhep{05}{2010}{013}{Analytic calculation of properties of holographic superconductors}.
\bibitem{hel1}
A.Donos and J.P.Gauntlett, \jhep{12}{2011}{091}{Holographic helical superconductors}.
\bibitem{hel2}
A.Donos and J.P.Gauntlett, \prl{108}{211601}{2012}{Helical superconducting black holes}.
\bibitem{cai15}
R.G.Cai, L.Li, L.F.Li and R.Q.Yang, {\it Introduction to Holographic Superconductor Models},
\hepth{1502.00437}{2015}.


\bibitem{pan11}
Q.Pan, J.Jing and B.Wang, \jhep{11}{2011}{088}{Analytical investigation of the phase transition between holographic insulator and superconductor in Gauss-Bonnet gravity}.
\bibitem{li11}
H.F.Li, R.G.Cai and H.Q.Zhang, \jhep{04}{2011}{028}{Analytical studies on holographic superconductors in Gauss-Bonnet gravity}.
\bibitem{gre09}
R.Gregory, S.Kanno and J.Soda, \jhep{10}{2009}{010}{Holographic superconductors with higher curvature corrections}.
\bibitem{cai10}
R.G.Cai, Z.Y.Nie and H.Q.Zhang, \prd{82}{066007}{2010}{Holographic p-wave superconductors from Gauss-Bonnet gravity}.
\bibitem{pan10}
Q.Pan, B.Wang, E.Papantonopoulos, J.Oliveira and A.Pavan, \prd{81}{106007}{2010}{Holographic superconductors with various condensates in Einstein-Gauss-Bonnet gravity}.
\bibitem{cai11back}
R.G.Cai, Z.Y.Nie and H.Q.Zhang, \prd{83}{066013}{2011}{Holographic phase transitions of p-wave superconductors in Gauss-Bonnet gravity with backreaction}.



\bibitem{zha15}
L.Zhang, Q.Pan and J.Jing, \plb{743}{104}{2015}{Holographic p-wave superconductor models with Weyl corrections}.
\bibitem{cha15}
P.Chaturvedi and G.Sengupta, \jhep{04}{2015}{001}{p-wave holographic superconductors from Born-Infeld black holes}.
\bibitem{zha13}
Z.Zhao, Q.Pan and J.Jing, \plb{719}{440}{2013}{ Holographic insulator/superconductor phase transition with Weyl corrections}.
\bibitem{jin12}
J.Jing, Q.Pan and S.Chen, \plb{716}{385}{2012}{Holographic superconductor/insulator transition with logarithmic electromagnetic field in Gauss Bonnet gravity}.


\bibitem{hor98}
G.T.Horowitz and R.C.Myers, \prd{59}{026005}{1998}{The AdS/CFT correspondence and a new positive energy conjecture for general relativity}.
\bibitem{nis10}
T.Nishioka, S.Ryu and T.Takayanagi, \jhep{03}{2010}{131}{Holographic superconductor/ insulator transition at zero temperature}.
\bibitem{wit98a}
E.Witten, \atmp{2}{505}{1998}{Anti-de Sitter space, thermal phase transition and confinement in gauge theories}.



\bibitem{alb08}
T.Albash and C.V.Johnson,  \jhep{09}{2008}{121}{A holographic superconductor in an external amgnetic field}.
\bibitem{ge10}
X.H.Ge, B.Wang, S.F.Wu, and G.H.Yang, \jhep{08}{2010}{108}{Analytical study on holographic superconductors in external magnetic field}.
\bibitem{ge12}
X.H.Ge and H.Q.Leng, \ptp{128}{2012}{1211}{Analytical Calculation on Critical Magnetic Field in Holographic Superconductor with Backreaction}.
\bibitem{cui13}
S.L.Cui and Z.Xue, \prd{88}{2013}{107501}{Critical magnetic field in a holographic superconductor in Gauss-Bonnet gravity with Born-Infeld electrodynamics}.
\bibitem{roy12}
D.Roychowdhury, \prd{86}{2012}{106009}{Effect of external magnetic field on holographic superconductors in presence of nonlinear corrections}.






\bibitem{alb09}
T.Albash and C.V.Johnson, \prd{80}{126009}{2009}{Vortex and droplet engineering in holographic superconductors}.
\bibitem{roy13}
D.Roychowdhury, \jhep{05}{2013}{162}{Holographic droplets in p-wave insulator/superconductor transition}.
\bibitem{cai13}
R.G.Cai, S.He, L.Li and L.F.Li, \jhep{12}{2013}{036}{A holographic study on vector condensate induced by a magnetic field}.
\bibitem{mae08}
K.Maeda and T.Okamura, \prd{78}{106006}{2008}{Characteristic length of an AdS/CFT superconductor}.
\bibitem{zen10a}
H.B.Zeng, Z.Y.Fan and H.S.Zong, \prd{81}{106001}{2010}{Superconducting coherence length and magnetic penetration depth of a p-wave holographic superconductor}.
\bibitem{zen10b}
H.B.Zeng, Y.Jiang, Z.Y.Fan and H.S.Zong, \prd{82}{126014}{2010}{Characteristic length of a holographic superconductor with d-wave gap}.
\bibitem{roy15}
D.Roychowdbury, \jhep{10}{2015}{018}{Chern-Simons vortices and holography}.
\bibitem{roy14}
D.Roychowdbury, {\it Towards holographic duals for anomalous supercurrents}, \hepth{1403.0085}{2014}
\bibitem{mae10}
K.Maeda, M. Natsuume and T.Okamura, \prd{81}{026002}{2010}{Vortex lattice for a holographic superconductor}.
\bibitem{mae11}
K.Maeda and T.Okamura, \prd{83}{066004}{2011}{Vortex flow for a holographic superconductor}.












\bibitem{shi12}
T.Shiromizu, S.Ohashi and R.Suzuki, \prd{86}{064041}{2012}{No-go on strictly stationary spacetimes in four/higher dimensions}.
\bibitem{bak13}
B.Bakon and M.Rogatko, \prd{87}{084065}{2013}{Complex scalar field in strictly stationary Einstein-Maxwell-axion-dilaton spacetime with negative cosmological constant}.

\bibitem{reg15}
M.Regis, J.Q.Xia, A.Cuoso, E.Branchini, N.Fornengo, and M.Viel, \prl{114}{241301}{2015}{Particle Dark Matter Searches Outside the Local Group}.
\bibitem{ali15}
Y.Ali-Haimoud, J.Chluba, and M.Kamionkowski, \prl{115}{071304}{2015}{Constrants on Dark Matter Interactions with Standard Model Particles from Cosmic Microwave Background Spectral Distortions}.



\bibitem{bra14}
J.Bramante and T.Linden, \prl{113}{191301}{2014}{Detecting dark matter with imploding pulsars in the galactic center}.
\bibitem{ful15}
J.Fuller and C.D.Ott, \mnras{450}{L71}{2015}{Dark-matter-induced collapse of neutron stars: a possible link between fast radio bursts and missing pulsar problem}.
\bibitem{lop14}
I.Lopes and J.Silk, \apj{786}{25}{2014}{A particle dark matter footprint on the first generation of stars}.
\bibitem{nak12}
A.Nakonieczna, M.Rogatko, and R.Moderski, \prd{86}{044043}{2012}{Dynamical collapse of charged scalar field in phantom gravity}.
\bibitem{nak15a}
A.Nakonieczna, M.Rogatko, and L.Nakonieczny, {\it Dark matter impact on gravitational collapse of an electrically charged scalar field}, \hepth{1508.02657}{2015}.



\bibitem{ger15}
A.Geringer-Sameth and M.G.Walker, \prl{115}{081101}{2015}{Indication of Gamma-Ray Emission from the Newly Discovered Dwarf Galaxy Reticulum II}.
\bibitem{bod15}
K.K.Boddy and J.Kumar, \prd{92}{023533}{2015}{Indirect detection of dark matter using MeV-range gamma-rays telescopes}.
\bibitem{til15}
K.Van Tilburg, N.Leefer, L.Bougas, and D.Budker, \prl{115}{011802}{2015}{Search for Ultralight Scalar Dark Matter with Atomic Spectroscopy}.

\bibitem{integral}
P.Jean {\it et al.}, \aa{407}{L55}{2003}{Early SPI/INTEGRAL measurements of 511 keV line emission from the 4th quadrant of the Galaxy}.
\bibitem{atic}
J.Chang {\it et al.}, \nat{456}{362}{2008}{An excess of cosmic ray electrons at energies of 300-800 GeV}.
\bibitem{pamela}
O.Adriani {\it et al.} (PAMELA Collaboration), \nat{458}{607}{2009}{An anomalous positron abundance in cosmic rays with energies 1.5-100 Gev}.
\bibitem{muon}
G.W.Bennett {\it et al.}, \prd{73}{072003}{2006}{Final report of the E821 muon anomalous magnetic moment measurement at BNL}.

\bibitem{massey15a}
D.Harvey, R.Massey, T.Kitching, A.Taylor and E.Tittley, \science{347}{1462}{2015}{The nongravitational interactions of dark matter in colliding galaxy clusters}.
\bibitem{massey15b}
R.Massey {\it et al.}, \mnras{449}{3393}{2015}{The behaviour of dark matter associated with four bright cluster galaxies in the 10 kpc core of Abell 3827}.


\bibitem{babar14}
J.P.Lees et al., \prl{113}{201801}{2014}{Search for a Dark Photon in $e^+ e^-$ Collisions at BABAR}.



\bibitem{nak14}
{\L}.Nakonieczny and M.Rogatko, \prd{90}{106004}{2014}{Analytic study on backreacting holographic superconductors with dark matter sector}.
\bibitem{nak15}
{\L}.Nakonieczny, M.Rogatko and K.I. Wysoki\'nski, \prd{91}{046007}{2015}{Magnetic field in holographic superconductors with dark matter sector}.


\bibitem{nak15a1} \L{}. Nakonieczny, M. Rogatko and K.I. Wysoki\'nski, 
{\it Analytic investigation of holographic phase transitions influenced by
dark matter sector}, \prd{92}{066008}{2015}.

\bibitem{rog15} M. Rogatko, K.I. Wysoki\'nski, {\it P-wave holographic superconductor/insulator phase
transitions affected by dark matter sector}, \hepth{1508.02869}{2015}.
\bibitem{pen15}
Y.Peng, {\it Holographic entangelment entropy in superconductor phase transition with dark matter sector}, \hepth{1507.07399}{2015}.

\bibitem{vac91}
T.Vachaspati and A.Achucarro, \prd{44}{3067}{1991}{Semilocal cosmic strings}.
\bibitem{ach00}
A.Achucarro and T.Vachaspati, \prep{327}{347}{2000}{Semilocal and electroweak strings}.
\bibitem{har09}
B.Hartmann and F.Arbabzadah, \jhep{07}{2009}{068}{Cosmic strings interacting with dark strings}.
\bibitem{bri09}
Y.Brihaye and B.Hartmann, \prd{80}{123502}{2009}{Effect of dark strings on semilocal strings}.
\bibitem{bri11} 
Y.Brihaye and B.Hartmann, \prd{83}{126008}{2011}{Holographic superfluid/fluid/insulator phase transitions in 2+1 dimensions}
\bibitem{dav12}
H.Davoudiasl, H.S.Lee and W.J.Marciano, \prd{85}{115019}{2012}{"Dark" Z implications for parity violation, rare meson decays, and Higgs physics}.
\bibitem{dav13}
H.Davoudiasl, H.S.Lee, I.Lewis and W.J.Marciano, \prd{88}{015022}{2013}{Higgs decays as a window into the dark sector}.


         


\bibitem{dome10} 
O. Domenech, M.Montull, A.Pomarol, A.Salvio and P.J. Silva, \jhep{08}{2010}{033}{Emergent gauge fields in holographic superconductors}.
\bibitem{mont12} 
M.Montull, O.Pujolas, A.Salvio, and P.J. Silva, \jhep{04}{2012}{135}{Magnetic response in the holographic insulator/superconductor transition}.



\end{thebibliography}
\end{document}